\documentclass[final,3p,times,twocolumn]{elsarticle}
 \biboptions{comma,sort&compress}
\usepackage{here}
 \usepackage{graphicx}
  \usepackage{epsfig}

\def\nin{\noindent}
\def\beq{\begin{equation}}
\def\eeq{\end{equation}}
\def\bea{\begin{eqnarray}}
\def\eea{\end{eqnarray}}
\def\nnb{\nonumber}

\journal{Nuc. Phys. (Proc. Suppl.)}

\begin{document}

\begin{frontmatter}



\title{Photon structure functions with heavy particle mass effects}
 \author[label1,label2]{Tsuneo Uematsu}
  \address[label1]{Graduate School of
Science, Kyoto University,
Kitashirakawa, Sakyo-ku,
Kyoto, 606-8502, Japan.}
\address[label2]{Maskawa Institute for Science and Culture, Kyoto Sangyo 
University, Kamigamo, Kita-ku, Kyoto 603-8555,Japan}
\ead{uematsu@scphys.kyoto-u.ac.jp}


\begin{abstract}
\noindent
In the framework of the perturbative QCD
we investigate heavy particle mass effects
on the unpolarized and polarized photon
structure functions, $F_2^\gamma$ and $g_1^\gamma$, respectively. We present
our basic formalism to treat heavy particle mass effects to NLO in perturbative
QCD.
We also study heavy quark effects on the QCD sum rule for the first moment of 
$g_1^\gamma$, which is related to axial anomaly. 
The photon structure function in supersymmetric QCD is also briefly discussed.

\end{abstract}

\begin{keyword}
QCD, photon structure function, heavy particle mass effects, sum rule, 
$g_1^\gamma$, axial anomaly


\end{keyword}

\end{frontmatter}


\section{Introduction}
\nin
In this talk I would like to address the question about how to incorporate the
heavy particle mass effects into the photon structure functions. This talk is
based on the works done in collaboration with Yoshio Kitadono, Ryo Sahara, Ken
Sasaki, Takahiro Ueda and Yutaka Yoshida.

Now let me make some remarks on why the photon structure is so interesting. 
First of all, 
it provides a good probe to study the QCD dynamics in perturbation
theory. The unpolarized virtual photon structure function $F_2^\gamma$ was
studied up to the next-to-next-to-leading order (NNLO) and the polarized
virtual photon structure function $g_1^\gamma$ was investigated 
to next-to-leading order (NLO).

The QCD sum rule for the first moment of $g_1^\gamma$ has attracted much 
attention in the literature since it is related to the axial anomaly.
In this talk we investigate the heavy particle mass effects on the photon
structure functions including the polarized structure function.

Here we investigate two-photon processes (Figure 1) with the kinematical region
where the mass squared of the probe photon ($Q^2$) is much larger than 
that of the target photon ($P^2$) which is in turn much bigger than the
$\Lambda_{\rm QCD}^2$, the QCD scale parameter squared. The advantage for 
studying
the virtual photon target is that we can calculate whole structure functions
up to next-leading-order (NLO), in contrast to the real photon target where
there remain uncalculable non-perturbative pieces.
\begin{figure}[hbt] 
\centerline{\includegraphics[width=5.cm]{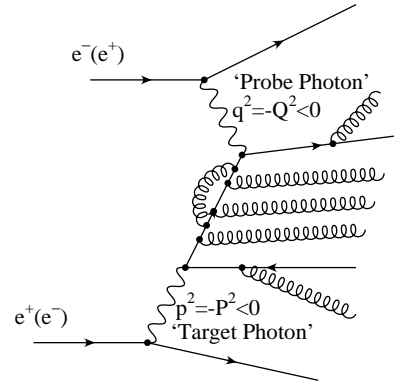}}
\caption{\scriptsize Deep-inelastic scattering on a virtual photon target
in the two-photon process of the e$^+$ e$^-$ collision experiments}
\label{fig1} 
\end{figure} 
\nin
\section{QCD calculation of photon structure functions}
\nin
The structure function $F_2^\gamma(x,Q^2)$ for the real photon target ($P^2=0$)
was first studied in the parton model \cite{Walsh-Zerwas, Kingsley} and then
studied to the leading order (LO) QCD based on the operator product expansion
(OPE) in ref.\cite{Witten}. 

The next-to-leading order QCD computation of $F_2^\gamma(x,Q^2)$ was performed
in the OPE and Renormalization Group (RG) method by Bardeen and Buras \cite{BB}
and in the QCD-improved parton model by Gl{\"u}ck and Reya \cite{GR} and
Fontannaz and Pilon \cite{FP} and by other people quoted in ref.\cite{SU}.

For a virtual photon target ($\Lambda_{\rm QCD}^2 \ll P^2 \ll Q^2$) 
the unpolarized
photon structure function $F_2^\gamma(x,Q^2,P^2)$ to LO in QCD was studied 
in ref.\cite{UWLO} and to NLO in ref.\cite{UWNLO}.

The master formula for the $n$-th moment to NLO is given by
\begin{eqnarray}
&&\hspace{-0.5cm}\int_0^1 dx x^{n-2}F_2^\gamma(x,Q^2,P^2)=  
\frac{\alpha}{4\pi}\frac{1}{2\beta_0} \times \nonumber\\
&&\hspace{-0.5cm}\Bigl[
\sum_{i=+,-,NS}
{\cal L}_i^n
\frac{4\pi}{\alpha_s(Q^2)}
\Bigl\{1-\left(\frac{\alpha_s(Q^2)}{\alpha_s(P^2)}\right)^{\lambda_i^n/2\beta_0
+1}\Bigr\}\nonumber\\
&&+\hspace{0.5cm}\sum_{i=+,-,NS}{\cal A}_i^n\Bigl\{1-\left(\frac{\alpha_s(Q^2)}
{\alpha_s(P^2)}\right)^{\lambda_i^n/2\beta_0}\Bigr\}\nonumber\\
&&+\hspace{0.5cm}\sum_{i=+,-,NS}{\cal B}_i^n\Bigl\{1-\left(\frac{\alpha_s(Q^2)}
{\alpha_s(P^2)}\right)^{\lambda_i^n/2\beta_0+1}\Bigr\}\nonumber\\
&&\vspace{2cm}\hspace{1.3cm}+\qquad {\cal C}^n +\quad {\cal O}(\alpha_s) \qquad
 \Bigr]~,
\label{master}
\end{eqnarray}
where ${\cal L}_i^n$, ${\cal A}_i^n$, ${\cal B}_i^n$ and ${\cal C}^n$
are computed from the 
one- and two-loop anomalous dimensions together with one-loop
coefficient functions.
All of them are shown to be renormalization-scheme independent.
$\alpha_s(Q^2)$ is the QCD running coupling constant,
$\beta_0$ is the one-loop beta function, and $\lambda_i^n \ (i=+,-,NS)$
denote the eigenvalues of one-loop anomalous dimensions
$\gamma_{ij}^{0,n}\ (i,j=\psi,G)$.

\section{Heavy quark mass effects}
\nin
We compute the deviation arising from heavy quark mass effects on the
photon matrix elements of twist-2 quark and gluon operators and 
the corresponding coefficient functions to NLO QCD \cite{KSUU1}. 

The moment of $F_2^\gamma$ can be decomposed as
\begin{eqnarray}
&&\hspace{-1cm}M_{2}^{\gamma}(n, Q^2, P^2,m^2)\equiv 
\int_{0}^{1}dx x^{n-2}  F_{2}^{\gamma}(x,Q^2,P^2,m^2)\nnb\\
&&\hspace{-0.7cm}= M_{2}^{\gamma}(n, Q^2, P^2,m^2=0)
+ \Delta M_{2}^{\gamma}(n, Q^2, P^2,m^2)~,
\end{eqnarray}
where the additional moment due to mass effects reads
\begin{eqnarray}
&& \hspace{-1cm}\Delta M_{2}^{\gamma}(n, Q^2, P^2, m^2) 
 = \int_{0}^{1}dx x^{n-2} \Delta F_{2}^{\gamma}(x,Q^2,P^2, m^2)\nnb\\ 
 &&\hspace{-1cm}= \frac{\alpha}{4\pi}\frac{1}{2\beta_{0}}
     \left[ \hspace{0.2cm}   
              \sum_{i=\pm,NS} \Delta \mathcal{A}_{i}^{n}
              \left[ 1 - \left( \frac{\alpha_s(Q^2)}
			        {\alpha_s(P^2)} \right)^{d_{i}^n } 
	      \right]
\right. \nnb\\
&& \left.\hspace{-1cm}
            + \sum_{i=\pm,NS} \Delta \mathcal{B}_{i}^{n}
              \left[ 1 - \left( \frac{\alpha_s(Q^2)}
			        {\alpha_s(P^2)} \right)^{d_{i}^n+1 } 
	      \right]
	    + \Delta \mathcal{C}^{n} 
\hspace{0.2cm} 
     \right] + \mathcal{O}(\alpha_s)~,\nnb 
\end{eqnarray}
where $d_i^n=\lambda_i^n/2\beta_0$  ($i=\pm,NS$).

In the massive quark limit: $\Lambda_{\rm QCD}^2\ll P^2\ll m^2\ll Q^2$
the deviations of the the coefficients turn out to be
\begin{eqnarray}
&&\hspace{-1cm}\Delta \mathcal{A}_{NS}^{n} 
 = -12 \beta_0 e_{H}^2 
(e_{H}^2 - \langle e^2 \rangle_{n_f})
     (\Delta \tilde{A}_{nG}^{\psi}/{n_f})~,
      \nnb\\
&&\hspace{-1cm}\Delta \mathcal{A}_{\pm}^{n} 
 = -12 \beta_0 e_{H}^2\langle e^2 \rangle_{n_f}
     (\Delta \tilde{A}_{nG}^{\psi}/{n_f})
     \frac{\gamma^{0,n}_{\psi\psi} - \lambda_{\mp}^{n} }
          { \lambda_{\pm}^{n} - \lambda_{\mp}^{n} }~,\nnb\\
&&\hspace{-1cm}\Delta \mathcal{B}_{NS}^{n} =
\Delta \mathcal{B}_{\pm}^{n} =0~,
  \quad
\Delta \mathcal{C}^{n}
 = 12 \beta_0 e_{H}^2 
     (\Delta \tilde{A}_{nG}^{\psi}/{n_f})~,\nnb
\end{eqnarray}
where $e_{H}$ is the heavy-quark charge and 
the average charge squared is defined as
$\langle e^2\rangle_{n_f}=\sum_{i=1}^{n_f}e_i^2/n_f$
with $n_f$ being the number of active flavors and the
deviation of operator matrix element 
$\Delta \tilde{A}_{nG}^{\psi}/{n_f}$ is
\begin{eqnarray}
 &&\hspace{-1cm}\Delta \tilde{A}_{nG}^{\psi}/{n_f}
 = 2\left[ - \frac{n^2+n+2}{n(n+1)(n+2)} 
               \ln \frac{m^2}{P^2}
	     + \frac{1}{n} - \frac{1}{n^2}
\right. \nnb \\
&& \left. \hspace{-1cm} \hspace{0.5cm}
	     + \frac{4}{(n+1)^2} - \frac{4}{(n+2)^2}
	     - \frac{n^2+n+2}{n(n+1)(n+2)} \sum_{j=1}^n\frac{1}{j}
       \right].\nnb\\
\end{eqnarray}
The Figure 2 shows the theoretical evaluation with charm quark mass
effects compared to the PLUTO's experimental data \cite{PLUTO} 
for the effective
structure function $F_{\rm eff}^\gamma(x,Q^2,P^2)=F_2^\gamma(x,Q^2,P^2)
+(3/2)F_L^\gamma(x,Q^2,P^2)$,
where $F_L^\gamma$ is the longitudinal structure function. 
The theoretical prediction (red curve)
with the charm quark mass effects shows the trend of reducing the massless
QCD calculation (purple curve) and becomes consistent with the experimental
data. Also shown is the theoretical curve (blue curve) with
the resummation prescription given in \cite{KSUU2}.
\begin{figure}[hbt] 
\centerline{\includegraphics[width=7.cm]{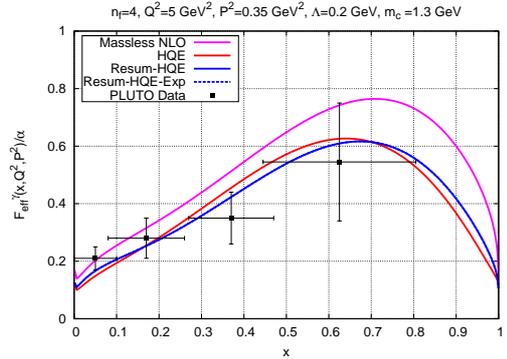}}
\caption{\scriptsize $F_{\rm eff}^\gamma$  to NLO in QCD with charm quark mass effects
 compared to PLUTO data for 
$n_f=4$, $Q^2=5$GeV$^2$, $P^2=0.35$GeV$^2$, $m_c=1.3$GeV \ \cite{KSUU2}}
\label{fig2} 
\end{figure} 
\nin
\section{Polarized photon structure function}
\nin
We now turn to the polarized photon structure function $g_1^\gamma$.
Let us first consider massless quark contributions to the photon
structure functions. 

For a real photon target ($P^2=0$), Bass, Brodsky and Schmidt
have shown that  
the 1st moment of $g_1^\gamma(x,Q^2)$
vanishes to all orders of $\alpha_s(Q^2)$ in QCD
\cite{BBS}:
\begin{equation}
\int_0^1 dx g_1^\gamma(x,Q^2)=0.
\end{equation}
Now the question is what about the $n=1$ moment of the virtual photon case.
By replacing the $x^{n-2}F_2^\gamma(x,Q^2,P^2)$ with 
$x^{n-1}g_1^\gamma(x,Q^2,P^2)$ in
(\ref{master}) and taking $n\rightarrow 1$ limit,
the first three terms vanish.
Denoting 
$e_i$, the $i$-th quark charge and $n_f$, the number of active flavors,
we have
\begin{equation}
\int_0^1 dx g_1^\gamma(x,Q^2,P^2)=
-\frac{3\alpha}{\pi}\sum_{i=1}^{n_f}{e_i}^4
+{\cal O}(\alpha\alpha_s)~.
\end{equation}
We can now go to ${\cal O}(\alpha\alpha_s)$ corrections
which turn out to be \cite{SU}:
\begin{eqnarray}
&&\hspace{-1cm}\int_0^1dx g_1^\gamma(x,Q^2,P^2)\nnb\\
&&\hspace{-0.5cm}=-\frac{3\alpha}{\pi}
\left[\sum_{i=1}^{n_f}e_i^4\left(1-\frac{\alpha_s(Q^2)}{\pi}\right)\right.
\nonumber\\
&&\left.\hspace{-0.5cm}-\frac{2}{\beta_0}(\sum_{i=1}^{n_f}e_i^2)^2\left(
\frac{\alpha_s(P^2)}{\pi}-\frac{\alpha_s(Q^2)}{\pi}\right)\right]
+{\cal O}(\alpha\alpha_s^2).\label{sum-rule-nlo}
\end{eqnarray}
This result coincides with the one obtained by Narison, Shore and 
Veneziano \cite{NSV}, apart from the overall sign for the definition
of $g_1^\gamma$.
We then extended this NLO result to the next-to-next-to-leading order (NNLO)
calculation, which is ${\cal O}(\alpha\alpha_s^2)$, where we 
need the 3-loop anomalous dimensions as well as 2-loop coefficient functions
 \cite{SUU}.

The same formalism can be applied to the heavy quark mass effects 
as in the case of unpolarized structure functions. Of course we have
to replace the one- and two-loop anomalous dimensions as well as the
one-loop coefficient functions by the polarized counterparts. In 
addition we put the operator matrix elements with the photon
states. Namely
\begin{eqnarray}
&&\hspace{-1cm} \Delta \tilde{A}_{nG}^{\psi} \frac{1}{n_f}
= 2\left[ - \frac{n-1}{n(n+1)} 
               \ln \frac{m^2}{P^2}
	     + \frac{1}{n} + \frac{1}{n^2}\right.\nnb\\
&&\left.\hspace{1.5cm} - \frac{4}{(n+1)^2} 
 - \frac{n-1}{n(n+1)} \sum_{j=1}^n\frac{1}{j}
       \right].
\end{eqnarray}

Let us first consider the heavy quark mass effects on the first moment
of the virtual photon structure function $g_1^\gamma(x,Q^2,P^2)$.
To the lowest order the first moment is given by
\bea
\int_0^1 dx g_1^\gamma(x,Q^2,P^2)=
\frac{\alpha}{4\pi}\frac{1}{2\beta_0}{\cal C}^{n=1}+{\cal O}
(\alpha\alpha_s).
\eea
For the massless case we have
\bea
\hspace{-0.7cm}
{\cal C}^{n=1}=12\beta_0\langle e^4\rangle_{n_f} 
(B_G^n+\widetilde{A}_{nG}^\psi)|_{n=1}
=-24\beta_0n_f\langle e^4\rangle_{n_f},
\eea
where $\langle e^4\rangle_{n_f}=\sum_{i=1}^{n_f}e_i^4/n_f$.
Now in the presence of the heavy quark, the $n_f$-th flavor, 
with the charge $e_{n_f}=e_H$:
\bea
\Delta{\cal C}^{n=1}=24\beta_0e_H^4~.
\eea
Hence we find that the heavy quark mass effects lead to
\bea
&&\hspace{-1cm}
\int_0^1 dx g_1^\gamma(x,Q^2,P^2)=
\frac{\alpha}{4\pi}\frac{1}{2\beta_0}
\left({\cal C}^{n=1}+\Delta{\cal C}^{n=1}\right)\nnb\\
&&\hspace{1.8cm}=-\frac{3\alpha}{\pi}\sum_{i=1}^{n_f-1}e_i^4+{\cal O}
(\alpha\alpha_s).
\eea
Thus the heavy quark decouples from the sum rule. This argument can
be extended to the order ${\cal O}(\alpha\alpha_s)$ and the sum rule
becomes eq.(\ref{sum-rule-nlo}) where $n_f$ is replaced by $n_f-1$.

We now investigate the heavy quark mass effects on the virtual
photon structure function $g_1^\gamma$ as a function of the Bjorken 
variable $x$ \cite{TU}. 
In Figure 3 we have plotted the theoretical prediction
for the structure function $g_1^\gamma(x,Q^2,P^2)$ in the case of
$Q^2=20$GeV$^2$ and $P^2=0.35$GeV$^2$, where $n_f=4$ and the heavy
quark is the charm quark with mass squared $m_c^2=2.25$GeV$^2$.
\begin{figure}[hbt] 
\centerline{\includegraphics[width=6.cm]{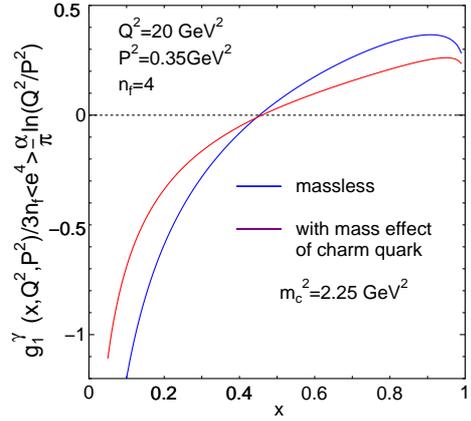}}
\caption{\scriptsize $g_1^\gamma(x,Q^2,P^2)$ 
in units of $3n_f\langle e^4\rangle(\alpha/\pi)\ln{Q^2/P^2}$
with and without
the charm quark mass effects for 
$n_f=4$, $Q^2=20$GeV$^2$, $P^2=0.35$GeV$^2$}
\label{fig3} 
\end{figure} 
\nin
\begin{figure}[hbt] 
\centerline{\includegraphics[width=6.cm]{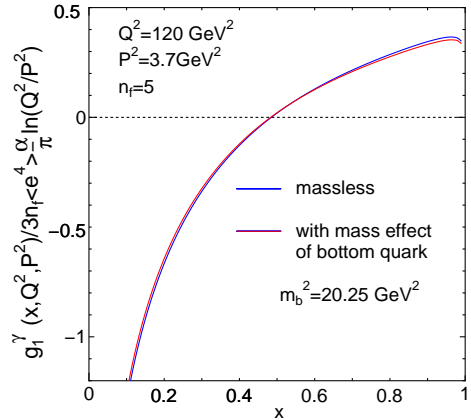}}
\caption{\scriptsize $g_1^\gamma(x,Q^2,P^2)$ 
in units of $3n_f\langle e^4\rangle(\alpha/\pi)\ln{Q^2/P^2}$
with and without
the bottom quark mass effects for 
$n_f=5$, $Q^2=120$GeV$^2$, $P^2=3.7$GeV$^2$}
\label{fig4} 
\end{figure} 
\nin

\nin
Note that the polarized structure function for the massive case 
(red curve) is suppressed in the large $x$ region compared to 
the massless case (blue curve) and becomes less negative for
small $x$ region when the charm quark mass effect is taken into 
account. Varying the values of $Q^2$ and $P^2$ for the kinematical
region where the charm quark can be regarded as the heavy quark we 
get the similar behaviors.

We have also shown in Figure 4, the structure function for 
the the case of $Q^2=120$GeV$^2$ and $P^2=3.7$GeV$^2$, where 
$n_f=5$ and the heavy quark is the bottom quark with mass squared 
$m_b^2=20.25$GeV$^2$. We note that the deviation from massless case
is almost negligible, since the charge factor is much smaller
compared to the charm case i.e. $e_b^4=(1/16)e_c^4$.

We have also applied our formalism for the photon structure
functions in supersymmetric QCD (SQCD) where there exist squarks and 
gluinos in addition to the quarks and gluons.
The squark contribution in the parton model was studied in \cite{KYSU}. 
For the SQCD radiative corrections,
we based our argument on the parton distribution functions (PDF's)
which satisfy the DGLAP-type evolution equation. The heavy mass
effects can be incorporated by the boundary conditions where we
require the PDF for each particle to vanish at $Q^2=m_i^2$ 
($i$=gluino, squark, heavy quark). By solving the coupled boundary
conditions we can determine the initial conditions and then we
get the solution for the moment of the each PDF. By performing
the inverse Mellin transform we finally get structure functions.
In Figure 5 we have plotted $F_2^\gamma(x,Q^2,P^2)$ as a function
of $x$. There we have also taken into account the threshold effects
$0 \leq x \leq x_{\rm max}$ with $x_{\rm max}=1/(1+P^2/Q^2+4m^2/Q^2)$.

\begin{figure}[hbt] 
\centerline{\includegraphics[width=8.cm]{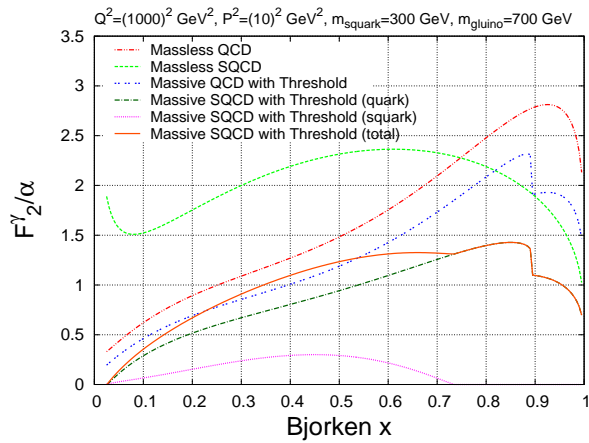}}
\caption{\scriptsize Photon structure function $F_2^\gamma(x,Q^2,P^2)/\alpha$
in supersymmetric QCD where heavy particle mass effects are taken into account
\cite{SUK}}
\label{fig4} 
\end{figure} 
\nin
We note that at small $x$
there is no significant difference between massless and massive
QCD, while there exists a large difference between massless and
massive SQCD. At large $x$, the significant mass-effects exist both
for non-SUSY and SUSY QCD. The SQCD case is seen to be much
suppressed at large $x$ compared to the QCD.

\section{Conclusion}
In this talk we discussed the heavy particle mass effects for the
virtual photon structure functions which
are calculable
by perturbative QCD (SQCD). Here we are particularly
interested in evaluation of unpolarized and polarized structure functions,
$F_2^\gamma$ and $g_1^\gamma$, with the heavy particle mass effects.
We first discussed our basic framework of studying heavy quark 
mass effects for $F_2^\gamma$. We then present the
analysis of the polarized photon structure function $g_1^\gamma$
in two aspects; QCD sum rule of the first moment which is related to the
axial anomaly as well as a the $x$ dependence of the structure function.
We have shown that the heavy particle decouples from the QCD sum rule. 
The analysis shows that the large $x$ region of 
$g_1^\gamma(x,Q^2,P^2)$ is suppressed in the presence of the heavy 
quark effects.
We also
investigated the photon structure functions in supersymmetric QCD. 
Experimental confrontation in the future is anticipated.

\section*{Acknowledgements}
\nin
I would like to thank Professor Stephen Narison 
for the wonderful organization and hospitality.





\begin{thebibliography}{999}
\vspace*{-0.25cm}
\bibitem{Walsh-Zerwas} T. F. Walsh and P. M. Zerwas, 
Phys. Lett. {\bf B44} (1973) 195.

\bibitem{Kingsley} R. L. Kingsley,
Nucl. Phys. {\bf B60} (1973) 45.

\bibitem{Witten} E. Witten,
Nucl. Phys. {\bf B120} (1977) 189.

\bibitem{BB} W. A. Bardeen and A. J. Buras, 
Phys. Rev. {\bf D20} (1979) 166.

\bibitem{GR} M. Gl{\"u}ck and E. Reya,
Phys. Rev. {\bf D28} (1983) 2749.

\bibitem{FP} M. Fontannaz and E. Pilon,
Phys. Rev. {\bf D45} (1992) 382.

\bibitem{SU} K. Sasaki and T. Uematsu,
Phys. Rev. {\bf D59} (1999) 114011.

\bibitem{UWLO} T. Uematsu and T. F. Walsh,
Phys. Lett. {\bf B101} (1981) 263.

\bibitem{UWNLO} T. Uematsu and T. F. Walsh,
Nucl. Phys.{\bf B199} (1982) 93.

\bibitem{KSUU1} Y. Kitadono, K.Sasaki, T. Ueda and T. Uematsu,
Prog. Theor. Phys. {\bf 121} (2009) 054019; Phys. Rev. {\bf D81}
(2010) 074029.

\bibitem{PLUTO} Ch. Berger et al.,
Phys. Lett. {\bf B142} (1984) 119.

\bibitem{KSUU2} Y. Kitadono, R. Sahara, T. Ueda and T. Uematsu,
Eur. Phys. J. {\bf C70} (2010) 999.


\bibitem{BBS} S. D. Bass, S. J. Brodsky and I. Schmidt,
Phys. Lett. {\bf B437} (1998) 417.

\bibitem{NSV} S. Narison,G. M. Shore and G. Veneziano, 
{Nucl. Phys.}\ {\bf B391} (1993) 69.

\bibitem{SUU} K. Sasaki, T. Ueda and T. Uematsu,
Phys. Rev. {\bf D73} (2006) 094024.


\bibitem{TU} T.~Uematsu, in preparation.


\bibitem{KYSU} Y. Kitadono, Y. Yoshida, R. Sahara and T. Uematsu,
Phys. Rev. {\bf D84} (2011) 074031.

\bibitem{SUK} R. Sahara, T. Uematsu and Y. Kitadono, 
Phys. Lett. {\bf B707} (2012) 517.

\end{thebibliography}








\end{document}